\documentclass[pacs]{revtex4}
\usepackage{amsmath}
\usepackage{amssymb}
\usepackage{graphicx}
\usepackage{color}
\begin{document}

\title{Thick brane solutions supported by two spinor fields}
\author{Vladimir Dzhunushaliev,$^{1,2}$}
\email{vdzhunus@krsu.edu.kg}
\author{Vladimir Folomeev,$^{2}$}
\email{vfolomeev@mail.ru}
\affiliation{
$^1$Institute for Basic Research,
Eurasian National University,
Astana, 010008, Kazakhstan; \\
$^2$Institute of Physicotechnical Problems and Material Science of the NAS
of the Kyrgyz Republic, 265 a, Chui Street, Bishkek, 720071,  Kyrgyz Republic
}


\begin{abstract}
Stationary thick brane solutions supported by two spinor fields are considered.
Two spinor fields are used here to exclude the
 off-diagonal components of the energy-momentum tensor of the spinor fields.
 The trapping of a test scalar field on the brane is also considered.
\end{abstract}


\maketitle

\section{Introduction}

At present, almost  all physical theories pretending to the role of a theory unifying
all known fundamental interactions are formulated  in spacetime having more than four dimensions.
Extra dimensions may have either microscopic size of order  of the fundamental Planck length or
they  may be large, and even infinite. In the first case, the smallness of extra  dimensions can be
achieved by introducing special mechanisms providing  compactification  of  extra dimensions
when  their  characteristic  size    becomes
much less than that of four-dimensional spacetime.

In models with large extra dimensions, it is assumed that
all matter fields are   confined on some hypersurface (called a brane)
which, in turn, is embedded in some multidimensional space
(called a bulk)  \cite{rs}. In such approach, two classes of branes are introduced: infinitely thin branes and  thick
branes. A brane is called infinitely thin if it has  a delta-like distribution of matter along the extra dimension.
However, from  a realistic point of view,  it is most 
reasonable to expect that 
 a brane has a finite thickness.
There are many works devoted to investigations in this direction (for a review, see e.g.  \cite{Dzhunushaliev:2009va}).
 All known thick brane solutions can be classified into two large groups: static solutions and solutions depending on time (cosmological thick branes).
From the mathematical  point of view, a search for
 thick brane solutions implies  the derivation of regular solutions of multidimensional field equations with a given material source.
 As a source of matter, various scalar fields are often used. The asymptotic geometry of spaceime outside a brane (the bulk geometry)
 can be either flat, de Sitter  or  anti-de Sitter.

 One of the main advantages of use of scalar fields in modeling thick branes is that on the one hand the models obtained
are comparatively simple, and on the other hand scalar fields are fundamental by their nature. However,
this does not exclude the possibility of creating brane models
in using other fundamental fields. For example, it could be spinor fields.
As we know, there are no so many solutions with spinor fields in general relativity. In particular,
 some cosmological solutions with spinor fields were found in Refs. \cite{Saha:2010zza}-\cite{Ribas:2010zj}.
In this connection, a search for other solutions would be interesting in itself.
As applied to branes, in  \cite{Dzhunushaliev:2009yx} we
considered a static model of a thick brane  supported by a nonlinear spinor field. In this model, the five-dimensional
regular solutions both with and without  cosmological term were found.

The solutions obtained in \cite{Dzhunushaliev:2009yx}
are completely static, when all physically measurable quantities (the metric function and the spinor field)
are to be time-independent. The next step is to examine a model when a spinor field is time-dependent. In this case,
since spinor field components are bilinear combinations such as $ \bar \psi \psi$,
one can seek a solution for the spinor field in the form
 $\psi (r,t)=f(r)\exp^{i\omega t}$,  where $f(r)$ is a column matrix which components depend only on the extra space coordinate
$r$. The presence of such
 time-dependent components of the spinor field leads to the appearance
 of the off-diagonal components of the energy-momentum tensor. To avoid this, we add to the model one more spinor field.
 In this case one can choose two spinor ans\"atze in such a way that the off-diagonal components of the energy-momentum tensor
 of every spinor field would have the opposite signs and annihilate each other.

The paper is organized as follows: in section \ref{gen_equations}
we derive
the field equations for the model of a thick brane  with two time-dependent spinor fields
having an arbitrary potential energy, and with a cosmological term taken into account. Choosing the special ans\"atze for these fields,
the off-diagonal components of the energy-momentum tensor are excluded.
In section \ref{special_poten} an example of numerical calculations for a special form of the potential energy is presented.
In section \ref{trapp_sect} the trapping of a test scalar field to the brane  is considered.

\section{Field equations}
\label{gen_equations}

We consider the five-dimensional gravitation with a nonlinear spinor field as a source of matter. The Lagrangian of the spinor field is
\begin{equation}
 \mathcal L_m = \frac{i}{2} \left(
		\bar \psi \not \! \nabla \psi - \bar \psi \overleftarrow{\not \! \nabla} \psi
	\right) - m \bar \psi \psi  + V(\bar \psi, \psi).
\label{2-10}
\end{equation}
We choose the potential $V(\bar \psi, \psi)$ satisfying the condition
\begin{equation*}
    \bar \psi \frac{\partial V}{\partial \bar \psi} = 2 V.
\end{equation*}
The corresponding five-dimensional Einstein and Dirac equations are
\begin{eqnarray}
  R_{ab} - \frac{1}{2} \eta_{ab} R &=& \varkappa T_{ab}
		+ \eta_{ab} \Lambda,
\label{2-20} \\
	\left[
	i \Gamma^a e_a^{\phantom{a} A}  D_A - m +
	\frac{\partial V}{\partial \bar \psi}
	\right] \psi &=& 0,
\label{2-30}
\end{eqnarray}
where $a=\bar 0, \bar 1, \bar 2, \bar 3, \bar 5$ is the Lorentz index;
$A=0,1,2,3,5$ is the world index; $e_a^{\phantom{a}A}$ is the 5-bein; $\Gamma^a$
are the five-dimensional Dirac matrices in a flat Minkowski space;
$D_A \psi = \left( \partial_A - \frac{1}{4} \omega_A^{\phantom{A} ab}
\Gamma_{ab} \right) \psi$ is the covariant derivative of the spinor $ \psi$;
$\Gamma_{ab} = \frac{1}{2}\left(\Gamma_a\Gamma_b-\Gamma_b\Gamma_a\right)$;
$ \not \! \nabla \psi=e^A_a \gamma^a D_A \psi$;
$m, \lambda$ are some parameters; $\Lambda$ is the cosmological constant; $\eta_{ab}$ is the five-dimensional
covariant Minkowski metric. The energy-momentum tensor for the spinor field is taken according to the textbook \cite{ortin}
\begin{equation}
 T_a^{\;A } = \frac{i}{2} \bar \psi \left(
		\Gamma^A e_a^{\phantom{a} B} + \Gamma_a g^{AB}
	\right) D_B \psi -
    \frac{i}{2} D_B \bar \psi \left(
		\Gamma^A e_a^{\phantom{a} B} + \Gamma_a g^{AB}
	\right) \psi -
    e_a^{\phantom{a} A} \mathcal L_m,
\label{2-40}
\end{equation}
where $\Gamma^A = e_a^{\,\, A} \Gamma^a$ are the
five-dimensional Dirac matrices in a curved spacetime;
$g^{AB} = e_a^{\phantom{a} A} e_b^{\,\, B} \eta^{ab}$ is the five-dimensional contravariant metric tensor;
$\bar \psi = \bar \psi^\dagger \Gamma^{\bar 0}$ is the Dirac conjugated spinor;
$D_A \bar \psi = \bar \psi\left ( \overleftarrow \partial_A + \frac{1}{4}
\omega_A^{\phantom{A} ab} \Gamma_{ab} \right)$ with
$\bar \psi \overleftarrow \partial_A = \partial_A \bar \psi$. Our definition of the energy-momentum tensor \eqref{2-40} has the opposite sign comparing with \cite{ortin} in order to be consistent with the definitions for $R_{ab}$ from \cite{poplawski}.

The five-dimensional Dirac matrices in a flat Minkowski space are
\begin{eqnarray}
  \Gamma^{\bar 0} &=& \begin{pmatrix}
		0												&		\mathbb I_{2 \times 2} \\
		\mathbb I_{2 \times 2} 	& 0
	\end{pmatrix},
\label{2-50}\\
  \Gamma^{\bar i} &=& \begin{pmatrix}
		0					&		-\sigma_{\bar i} \\
		\sigma_{\bar i}  & 0
	\end{pmatrix}, \bar i = 1,2,3,
\label{2-60} \\
  \Gamma^{\bar 5} &=& \begin{pmatrix}
		-i \mathbb I_{2 \times 2}	&		0 \\
		0   											& i \mathbb I_{2 \times 2}
	\end{pmatrix},
\label{2-70}
\end{eqnarray}
where $\mathbb I_{2 \times 2}$ is $2 \times 2$ unity matrix, and $\sigma_{\bar i}$
are Pauli matrixes
$$
  \sigma_{\bar 1} = \begin{pmatrix}
		0	 &		1 \\
		1  & 0
	\end{pmatrix}, \quad
\sigma_{\bar 2} = \begin{pmatrix}
		0	 & -i \\
		i  & 0
	\end{pmatrix}, \quad
\sigma_{\bar 3} = \begin{pmatrix}
		1	 &	0 \\
		0  & -1
	\end{pmatrix}.
$$
We seek a wall-like solution of the system \eqref{2-20}-\eqref{2-30}. To do this let us choose the five-dimensional bulk metric in the form
\begin{equation}
    ds^2 = \phi^2 (r) \left(
        e^{2 \chi(r)} dt^2 -dx^2 - dy^2 -dz^2
    \right) - dr^2.
\label{2-90}
\end{equation}

Next, we are going to consider a system consisting of two spinor fields
 $\psi_{1,2}$.
The reason for this is to exclude off-diagonal components of the energy-momentum tensor.
For the spinor fields, we use the following time-dependent ans\" atze
\begin{equation}
 \psi_1 = e^{i \omega t} \begin{pmatrix}
		a(r) 	\\
		0 		\\
		b(r)	\\
		0
	\end{pmatrix}
    \quad \text{and} \quad
 \psi_2 = e^{i \omega t} \begin{pmatrix}
		0 	   \\
		a(r)   \\
		0	   \\
		b(r)
	\end{pmatrix}.
\label{2-110}
\end{equation}
Then the corresponding components of the energy-momentum tensor \eqref{2-40}  are
\begin{eqnarray}
  (T_1)_{\bar 0 \bar 0} &=& (T_2)_{\bar 0 \bar 0} =
  -\frac{\omega e^{-\chi}}{\phi}
  \left(a^2 + b^2 \right) + V(\bar \psi, \psi),
\label{2-120}\\
  (T_1)_{\bar 0 \bar 3} &=& - (T_2)_{\bar 0 \bar 3} =
  \frac{\omega e^{-\chi}}{\phi}
  \left(a^2 - b^2 \right) + \frac{1}{2} ab\chi' ,
\label{2-130}\\
  (T_1)_{\bar 1 \bar 1} &=& (T_2)_{\bar 1 \bar 1} =
  (T_1)_{\bar 2 \bar 2} = (T_2)_{\bar 2 \bar 2} =
  (T_1)_{\bar 3 \bar 3} = (T_2)_{\bar 3 \bar 3} =
  V(\bar \psi, \psi),
\label{2-140}\\
  (T_1)_{\bar 5 \bar 5} &=& (T_2)_{\bar 5 \bar 5} =
  a b' - a'b + V(\bar \psi, \psi).
\label{2-150}
\end{eqnarray}
It is easily seen from expressions \eqref{2-130} that
 {\it the contributions from $\psi_1$ and $\psi_2$ have the opposite signs for the off-diagonal components $T_{\bar 0 \bar3}$.} Consequently,
 the combined energy-momentum tensor for two spinor fields has the diagonal components only. This allows
 the possibility of writing the  Einstein-Dirac equations in the form
\begin{eqnarray}
  \frac{\phi''}{\phi} + \frac{{\phi'}^2}{\phi^2} &=& \frac{2 \varkappa}{3} \left[
    \frac{\omega e^{-\chi}}{\phi} \left(a^2 + b^2 \right) - V
  \right] - \frac{\Lambda}{3} ,
\label{2-160}\\
  \frac{\phi''}{\phi} + \frac{{\phi'}^2}{\phi^2} + \frac{\chi''}{3} +
  \frac{{\chi'}^2}{3} + \frac{4}{3} \frac{\phi' \chi'}{\phi} &=&
  - \frac{2 \varkappa}{3} V - \frac{\Lambda}{3} ,
\label{2-170}\\
  \frac{{\phi'}^2}{\phi^2} + \frac{1}{2} \frac{\phi' \chi'}{\phi} &=&
  \frac{\varkappa}{3} \left(
    ab' -a'b - V
  \right) - \frac{\Lambda}{6} ,
\label{2-180}\\
  a' - \frac{\omega}{\phi} e^{-\chi} b+ a \left(
    \frac{2 \phi'}{\phi} + \frac{\chi'}{2} - m
  \right) + \frac{\partial V}{\partial \bar \psi} &=& 0,
\label{2-190}\\
  b' + \frac{\omega}{\phi} e^{-\chi} a + b \left(
    \frac{2 \phi'}{\phi} + \frac{\chi'}{2} + m
  \right) - \frac{\partial V}{\partial \bar \psi} &=& 0.
\label{2-200}
\end{eqnarray}
Using the redefinitions $r/\sqrt \varkappa \rightarrow r$,
$\omega \sqrt \varkappa \rightarrow \omega$,
$\lambda / \varkappa \rightarrow \lambda$,
$\varkappa^3 a^4 \rightarrow a^4$, $\varkappa^3 b^4 \rightarrow b^4$,
$\varkappa \Lambda \rightarrow \Lambda$, $m \sqrt \varkappa \rightarrow m$ and
performing
some manipulations with equations \eqref{2-160}-\eqref{2-200}, we obtain the following set of equations:
\begin{eqnarray}
  \frac{\phi''}{\phi} + \frac{{\phi'}^2}{\phi^2} &=& \frac{2}{3} \left[
    \frac{\omega e^{-\chi}}{\phi} \left(a^2 + b^2 \right) - V
  \right] - \frac{\Lambda}{3} ,
\label{2-210}\\
  \chi'' + {\chi'}^2 + 4 \frac{\phi' \chi'}{\phi} &=&
  - 2 \frac{\omega e^{-\chi}}{\phi} \left(a^2 + b^2 \right) ,
\label{2-220}\\
  \frac{{\phi'}^2}{\phi^2} + \frac{1}{2} \frac{\phi' \chi'}{\phi} &=&
  - \frac{1}{3} \left[
    \frac{\omega e^{-\chi}}{\phi} \left(a^2 - b^2 \right) + 2 m ab - V
  \right] - \frac{\Lambda}{6} ,
\label{2-230}\\
  a' - \frac{\omega}{\phi} e^{-\chi} b+ a \left(
    \frac{2 \phi'}{\phi} + \frac{\chi'}{2} - m
  \right) + \frac{\partial V}{\partial \bar \psi} &=& 0,
\label{2-240}\\
  b' + \frac{\omega}{\phi} e^{-\chi} a + b \left(
    \frac{2 \phi'}{\phi} + \frac{\chi'}{2} + m
  \right) - \frac{\partial V}{\partial \bar \psi} &=& 0.
\label{2-250}
\end{eqnarray}
From equation \eqref{2-230}, we see the following constraints on the boundary conditions
\begin{equation}
\label{2-260}
    \frac{\omega e^{-\chi_0}}{\phi_0} \left(a_0^2 - b_0^2 \right) + 2 m a_0b_0 - V_0
    + \frac{\Lambda}{2} = 0,
\end{equation}
where the index zero refers to the quantities on the brane, i.e. when $r=0$.

The spinors \eqref{2-110} are chosen in the spinor representation, and in the standard representation the spinor components
have to be either odd or even functions. Using this fact,
henceforth we will assume that
$a_0 = b_0$.

\section{Numerical calculations for a special potential $V(\bar \psi, \psi)$
}
\label{special_poten}

In this section we consider a case when the potential $V(\bar \psi, \psi)$ from \eqref{2-10} is chosen in the form
\begin{equation}
\label{spec_pot}
V = \frac{\lambda}{2} \left( \bar \psi \Gamma^{[A} \Gamma^B \Gamma^{C]} \psi \right)
\left( \bar \psi \Gamma_{[A} \Gamma_B \Gamma_{C]} \right) \psi,
\end{equation}
where $[ \cdots ]$ denotes the antisymmetrization. For the potential given by \eqref{spec_pot}, we have
\begin{eqnarray}
  V\left( \bar \psi_{1,2}, \psi_{1,2} \right) &=& - 4 \lambda a^2 b^2,
\label{3-10}\\
  \frac{\partial V}{\partial \bar \psi_1} &=& \lambda \left(
  \Gamma^{[A} \Gamma^B \Gamma^{C]} \psi_1 \right)
  \left( \bar \psi_1 \Gamma_{[A} \Gamma_B \Gamma_{C]} \psi_1 \right) =
  - 4 \lambda e^{i \omega t}
  \begin{pmatrix}
		a^2 b 	\\
		0 		\\
		a b^2	\\
		0
	\end{pmatrix},
\label{3-20}\\
  \frac{\partial V}{\partial \bar \psi_2} &=&
  - 4 \lambda e^{i \omega t}
  \begin{pmatrix}
		0 	        \\
		a^2 b 		\\
		0	        \\
		a b^2
    \end{pmatrix}.
\label{3-30}
\end{eqnarray}
Using this potential, we can derive the following system of equations:
\begin{eqnarray}
  \frac{\phi''}{\phi} + \frac{{\phi'}^2}{\phi^2} &=& \frac{2}{3} \left[
    \frac{\omega e^{-\chi}}{\phi} \left(a^2 + b^2 \right) + 4 \lambda a^2 b^2
  \right] - \frac{\Lambda}{3} ,
\label{3-40}\\
  \chi'' + {\chi'}^2 + 4 \frac{\phi' \chi'}{\phi} &=&
  - 2 \frac{\omega e^{-\chi}}{\phi} \left(a^2 + b^2 \right) ,
\label{3-50}\\
  \frac{{\phi'}^2}{\phi^2} + \frac{1}{2} \frac{\phi' \chi'}{\phi} &=&
  - \frac{1}{3} \left[
    \frac{\omega e^{-\chi}}{\phi} \left(a^2 - b^2 \right) + 2 m ab + 4 \lambda a^2 b^2
  \right] - \frac{\Lambda}{6} ,
\label{3-60}\\
  a' - \frac{\omega}{\phi} e^{-\chi} b+ a \left(
    \frac{2 \phi'}{\phi} + \frac{\chi'}{2} - m
  \right) - 4 \lambda a^2 b &=& 0,
\label{3-70}\\
  b' + \frac{\omega}{\phi} e^{-\chi} a + b \left(
    \frac{2 \phi'}{\phi} + \frac{\chi'}{2} + m
  \right) + 4 \lambda a b^2 &=& 0
\label{3-80}
\end{eqnarray}
with the boundary conditions
\begin{equation}
\label{3-90}
   a_0 = b_0, \quad \phi_0 = 1, \quad {\phi_0}' = 0, \quad
   \chi_0 = 0, \quad {\chi_0}' = 0
\end{equation}
and the constraint
\begin{equation}
\label{3-100}
   \Lambda = -4 a_0^2 \left( m + 2 \lambda a_0^2 \right)
\end{equation}
which follows from equation \eqref{3-60}.

\begin{figure}[t]
\begin{minipage}[t]{.49\linewidth}
  \begin{center}
  \includegraphics[width=9.5cm]{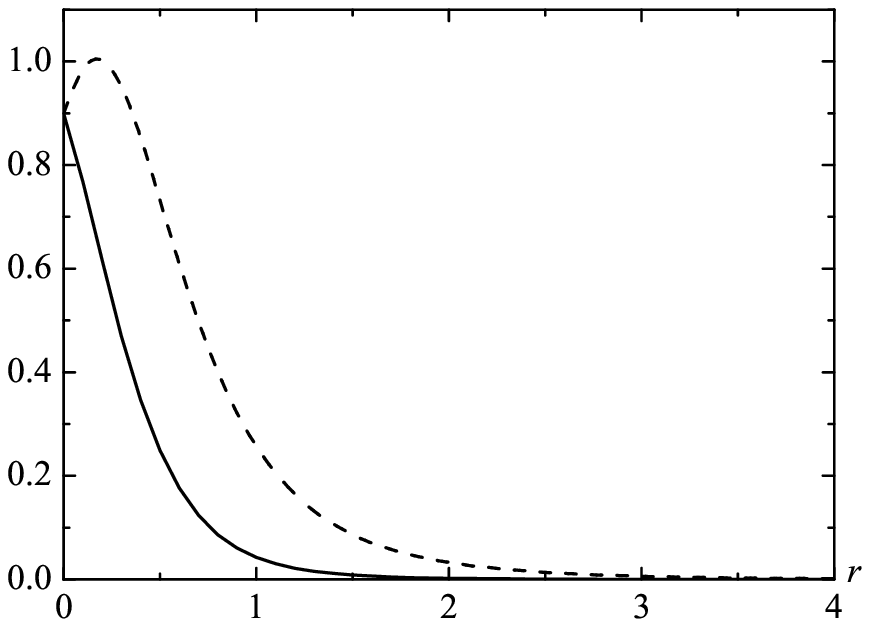}
\vspace{-1.cm}
  \caption{\small The profiles of the eigenfunctions $a(r)$ (the solid line) and $b(r)$
  (the dashed line).}
  \label{fig1}
  \end{center}
\end{minipage}\hfill
\begin{minipage}[t]{.49\linewidth}
  \begin{center}
  \includegraphics[width=9.2cm]{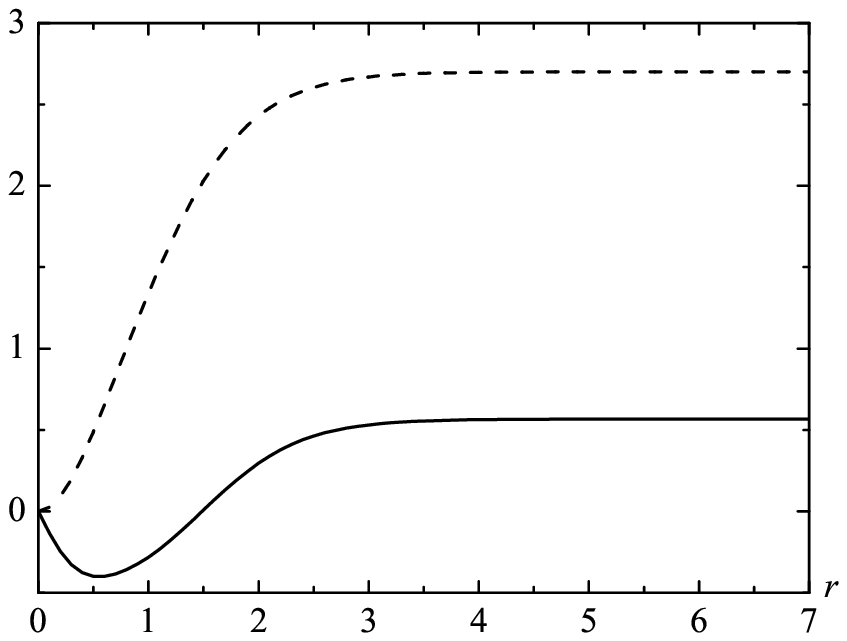}
\vspace{-1.3cm}
  \caption{\small The profiles of $\phi'(r)/\phi(r)$ and $\chi(r)$ are shown by the solid and dashed lines, respectively.}
  \label{fig2}
  \end{center}
\end{minipage}\hfill
\end{figure}

The numerical analysis shows that regular solutions of the set of equations
\eqref{3-40}-\eqref{3-80} do exist only for an appropriate choice of the parameter $\omega$.
One can say that
this system of equations describes
 a nonlinear eigenvalue problem for the eigenfunctions $a(r), b(r)$.
Then the statement of the problem is as follows:
Given some values  of  $a_0, m$ and $\lambda$ which enter into equations \eqref{3-40}-\eqref{3-80}
and  boundary  conditions \eqref{3-90}, find the corresponding eigenvalue of  $\omega$.
 We  solved  the  system  numerically  using  the
NDSolve routine from {\it Mathematica}.
As an example we used the following values of the parameters: $a_0=0.9, m=1.0, \lambda=-0.25$
which give us $\omega \simeq -1.45227015$.
 The results of the numerical calculations are shown
 in Figs. \ref{fig1} and \ref{fig2}.

\section{Trapping of matter}
\label{trapp_sect}

The five-dimensional localized wall-like solutions obtained above can be used for a description of a brane only
if it will be possible to show that various test matter  fields can be confined on such wall.
As an example of such field, let us consider here a test complex scalar filed $\eta$ with the Lagrangian
$$
	L_{\eta}=\frac{1}{2}\partial_A \eta^{*}\partial^A\eta-
	\frac{1}{2}m_0^2 \eta^*\eta,
$$
where $m_0$ is the mass of the test field. Using this Lagrangian, we find the following equation for the scalar field
\begin{equation}
\label{tf_eq}
	\frac{1}{\sqrt{-^5 g}}\frac{\partial}{\partial x^A}\left(
		\sqrt{-^5 g} g^{AB}\frac{\partial \eta}{\partial x^B}
	\right)=-m_0^2 \eta.
\end{equation}
Here $\sqrt{-^5 g}$ is the determinant of the five-dimensional
metric $g_{AB}$, and $\eta$ is a function of all coordinates $\eta=\eta(x^A)$. Taking into account that the  canonically conjugate momenta $p_\mu=(E, \overrightarrow{p})$ are integrals of motion, we seek a solution in the form
$$
  \eta (x^{A}) = X(r) \exp (-ip_{\mu }x^{\mu }).
$$
Substituting this ansatz in  \eqref{tf_eq}, one can find the following equation for $X(r)$
$$
    X^{\prime\prime} + \left(4\frac{\phi^{\prime}}{\phi}+\chi^\prime\right)X^{\prime}+ (p^{\mu }p_{\mu } -m_{0}^2 ) X = 0,
$$
or, taking into account that
$p^{\mu}p_{\mu}= \phi^{-2} \left(e^{-2\chi}E^2-\overrightarrow{p}^2\right)$, we have
\begin{equation}
\label{test_eq}
	X^{\prime\prime} +\left(4\frac{\phi^{\prime}}{\phi}+\chi^\prime\right)X^{\prime}+ \left[
	\left(e^{-2\chi}E^2-\overrightarrow{p}^2\right)\phi^{-2}-m_0^2
	\right] X = 0,
\end{equation}
where  the prime denotes differentiation with respect to $r$.
To find a solution of this equation, let us determine the asymptotic behavior of the metric functions $\phi$ and $\chi$.
Taking into account the numerical results obtained in the previous section, one can see that asymptotically, as
$r \to \infty$, $\chi \to const$.
Then, using equation \eqref{3-60}, one can easily find that $\phi \sim \exp(\sqrt{-\Lambda/6}r)$.
Obviously, this expression is only valid for
 negative $\Lambda$. Correspondingly, equation \eqref{test_eq} takes the following asymptotic form
$$
		X^{\prime\prime}+4\sqrt{-\frac{\Lambda}{6}}X^{\prime}-m_0^2  X = 0,
$$
with the asymptotically decaying solution
\begin{equation}
\label{asymp_sol_minus}
	X_{\infty}\simeq D \exp{\left[-2\left(\sqrt{-\frac{\Lambda}{6}}+
\sqrt{-\frac{\Lambda}{6}+\frac{m_0^2}{4}}\,\,\right)|r|\right]},
\end{equation}
where $D$ is an integration constant.
One can see from this solution that the test scalar field decreases
exponentially fast
 that corresponds to the fact that this field is concentrated around the brane.

As a necessary condition  for the trapping of matter on the brane,
one can require converging the field energy per unit 3-volume of the brane \cite{Abdyrakhmanov:2005fs}, i.e.,
\begin{equation}
\label{E_tot}
     E_{\rm tot}[\eta] = \int \limits_{-\infty}^{\infty} T^0_0 \sqrt{-^5g} \,d r
          = \int\limits_{-\infty}^{\infty}
            \phi^4 e^{\chi}\left[ \frac{1}{\phi^2}(e^{-2\chi}E^2 + \overrightarrow{p}^2)X^2
	    + m_0^2 X^2 + X^{\prime 2} \right] d r < \infty,
\end{equation}
and also the norm of the field $\eta$ should be finite
$$
	||\eta||^2 = \int\limits_{-\infty }^{\infty }\sqrt{-^5g}\,\eta^*\eta\,d r
    	       = \int_{-\infty }^{\infty } \phi^4 e^{\chi}\,X^2 \,d r.
$$
Taking into account the asymptotic solution \eqref{asymp_sol_minus}, one can see that both $E_{\rm tot}$ and $||\eta||$ converge asymptotically.
Then it becomes obvious from the above analysis that the localized solutions obtained in section \ref{special_poten} confine the test scalar field,
and this indicates that such solutions can be interpreted as brane solutions.


Summarizing the results, we have obtained  the $Z_2$-symmetric thick brane stationary solutions supported by two
nonlinear  spinor fields in the presence of  the five-dimensional cosmological $\Lambda$-term. We have shown
that such solutions do exist for the special spinor ans\"atze  given by \eqref{2-110}. This ans\"atze allowed us to exclude
 the off-diagonal components of the energy-momentum tensor. It happens because the currents for the spinor fields are
\begin{eqnarray}
\label{dis-10}
  \left( J_{\psi_1} \right)^0 &=& \left( J_{\psi_2} \right)^0 =
  \frac{a^2 + b^2}{\phi}, \\
\label{dis-20}
  \left( J_{\psi_1} \right)^3 &=& - \left( J_{\psi_2} \right)^3 =
  \frac{a^2 - b^2}{\phi},
\end{eqnarray}
and consequently
\begin{eqnarray}
\label{dis-30}
  J^0 &=& \left( J_{\psi_1} \right)^0 + \left( J_{\psi_2} \right)^0 =
  2 \left( \frac{a^2 + b^2}{\phi} \right), \\
\label{dis-40}
  J^3 &=& \left( J_{\psi_1} \right)^3 + \left( J_{\psi_2} \right)^3 = 0.
\end{eqnarray}
This means that the current $J^A$ along the $z$ axis for the case of two spinor fields is equal to zero.
This in turn leads to the vanishing of the off-diagonal components of the energy-momentum tensor.

Using the ans\"atze \eqref{2-110}, we performed the numerical calculations of the set of equations \eqref{3-40}-\eqref{3-80}
with a special choice of the potential $V(\bar \psi, \psi)$ in the form \eqref{spec_pot}.
Equations \eqref{3-40}-\eqref{3-80} constitute an eigenvalue problem for $\omega$ subject to  boundary conditions
\eqref{3-90} and constraint \eqref{3-100}. The example of  regular solutions of the  above system
is shown in Figs. \ref{fig1} and \ref{fig2}. The asymptotic behavior of the obtained solutions corresponds to  an anti-de Sitter spacetime
($\Lambda<0$). Using the asymptotic solutions,  in section \ref{trapp_sect} it was shown that they can trap
the test scalar field.    It indicates that such solutions may be interpreted as brane solutions.

\section*{Acknowledgements}

We are grateful to the Research Group Linkage Programme of the Alexander von Humboldt Foundation for the support of this research.

\end{document}